\documentclass[runningheads]{llncs}
\usepackage{graphicx}
\usepackage{cite}

\usepackage{amsmath,amssymb,amsfonts}
\usepackage{nameref}
\usepackage{varioref}
\usepackage{hyperref}
\usepackage{cleveref}
\usepackage{graphicx}
\usepackage{textcomp}
\usepackage{xcolor}
\usepackage{multirow}
\usepackage[T1]{fontenc}
\usepackage{fourier, erewhon}
\usepackage{array, caption, floatrow, tabularx, makecell, booktabs, subcaption}%
\setcellgapes{3pt}
\def\BibTeX{{\rm B\kern-.05em{\sc i\kern-.025em b}\kern-.08em
    T\kern-.1667em\lower.7ex\hbox{E}\kern-.125emX}}
    
\usepackage{fancyhdr}
\usepackage{lipsum}
\usepackage{algorithm}
\usepackage{wrapfig}
\usepackage{algorithmic}
\usepackage{multicol}
\usepackage[title]{appendix}
\usepackage[tableposition=top]{caption}
\usepackage{floatrow}
\DeclareMathOperator*{\argmax}{arg\,max}

\usepackage{indentfirst}
\begin{document}

\title{EfficientRec: An unlimited user scale recommendation system based on clustering and user’s interaction embedding profile
}

\author{Vu Hong Quan\inst{1} \and
Le Hoang Ngan\inst{1} \and
Le Minh Duc\inst{1} \and
Nguyen Tran Ngoc Linh\inst{1} \and
Hoang Quynh - Le\inst{2}}
\authorrunning{Vu Hong Quan et al.}
%
\institute{{Data Analytics Center - Viettel Telecom} - {Viettel Group} \and
{University of Engineering and Technology} - {Vietnam National University} \\
\email{\{quanvh8, nganlh, duclm29, linhntn3\}@viettel.com.vn; lhquynh@vnu.edu.vn}}
\maketitle              


\begin{abstract}
Recommendation systems are highly interested in technology companies nowadays. The businesses are constantly growing users and products, causing the number of users and items to continuously increase over time, to very large numbers. Traditional recommendation algorithms with complexity dependent on the number of users and items make them difficult to adapt to the industrial environment. In this paper, we introduce a new method applying graph neural networks with a contrastive learning framework in extracting user preferences. We incorporate a soft clustering architecture that significantly reduces the computational cost of the inference process. Experiments show that the model is able to learn user preferences with low computational cost in both training and prediction phases. At the same time, the model gives a very good accuracy. We call this architecture EfficientRec\footnote{Our source code available at: https://github.com/quanvu0996/EfficientRec} with the implication of model compactness and the ability to scale to unlimited users and products.


\end{abstract}


\section{Introduction}

\par Personalization is the topic of investment with high returns in recent years. Two typical collaborative filtering (CF) algorithms for the recommendation problem are matrix factorization \cite{b37} and two-head DNNs (\cite{b34}, \cite{b35}). While recent studies focus on the accuracy of the recommendation system in the lab and achieve positive results such as BiVAE \cite{b32}, VASP \cite{b33},... We find these methods facing difficulties in deploying on the production environment because:
\par First, the architecture of these models is not optimized. These models use user\_id embedding as part of the model. This leads to the calculation complexity and size of the model depending on the number of users. When the number of users increases, it is forced to re-train a new model, the model cannot recommend for new users. When the number of users is large, the size and calculation complexity of the model become very large (\cite{b2}, \cite{b5}). These limitations in reality are very common in businesses when the number of users is large and constantly increasing.
\par Second, the item selection process of these models is not optimal. Models deployed in the lab often try to predict the entire utility matrix. The reason is to evaluate a model's performance in the laboratory, we must rely on its predictions for a few items that have been rated in the test set. If a user is recommended items that the user has not rated, we do not know whether the recommendation is correct or not. Predicting the entire utility matrix is very expensive, for each user we have to calculate for both items that are liked and not liked. Unlike the lab, the production recommendations will be evaluated on A/B Testing. A product that has not been rated in the past if it is recommended, users can interact and show their level of favor. The production recommendation is to search for the items that the user liked.

\par Some papers try to solve the first problem by using online learning \cite{b6} or parallel ( \cite{b7}, \cite{b8} ) matrix factorization. More recently, \cite{b9} proposed a distributed alternative stochastic gradient distribution solver for an LFA-based recommendation based on a distribution mechanism including efficient data partitioning. Clustering is a potential idea to solve both problems (\cite{b20}, \cite{b21}, \cite{b22}). Clustering could improve recommendation accuracy \cite{b18}, increase the diversity of lists of recommendations \cite{b19},...Although these methods help the model become trainable on a large scale, the model size and cold start problems still remain.

\par In this paper, we try to solve these two non-optimal points, and propose a new architecture that can effectively be implemented in the production environment. Our main contributions include:
\begin{itemize}
\item Building recommendation system with complexity independent of the number of users. It does not use user\_id so it can scale unlimited with the number of users without affecting the performance of the model. At the same time, the model can operate with new users without having to build and retrain the model.
\item Proposing the algorithm of clustering item selection to help prune a large number of unnecessary parts of the utility matrix, making the personalization problem become search items that users liked.
\item Application of contrastive learning architecture to extract the user's preference effectively.
\end{itemize}

\section{Related works}

\subsection{Graph Neural Network in Recommend System}
In recent years, studies on the application of GNN to recommendation systems have been proposed. The most intuitive reason is that GNN techniques have demonstrated to power representational learning for graph data in a variety of domains. User-item interaction prediction  is one of classic problems in recommendation, then, user-item interaction data can be represented by a bipartite graph, the edges corresponding to the interaction of the user-item. Van Den Berg et.al \cite{b24} first applied GNN with GC-MC model on user-item rating graph to learn embedded representation of user and item. In fact, the recommended dataset can be up to billions of nodes and edges, where each node contains many features, it is difficult to apply traditional GNN models due to large memory usage and long training time. To deal with the large-scale graphs, one of the classic ways is to apply graph sampling. GraphSAGE \cite{b25} randomly samples a fixed number of neighbors, and PinSage \cite{b26} employs the random walk strategy for sampling. However, sampling will lose more or less part of the information, and few studies focus on how to design an effective sampling strategy to balance the effectiveness and scalability.

\subsection{Self-supervised learning}
Self-supervised learning (SSL) as a technique to learn with unlabeled data, recently applied to recommendation for mitigating the data sparsity issue. The basic idea is to augment the training data with various data augmentation, and supervised tasks to predict or reconstitute the original examples as auxiliary tasks. SSL has been widely applied in many fields, such as data augmentation methods in image processing or masked language tasks in BERT \cite{b30} model applied in natural language processing. Inspired by the success of SSL in other fields, recent studies have applied SSL to recommendation systems and have achieved remarkable achievements. Kun Zhou et.al \cite{b27} introduced the $S^3-Rec$ model, the main idea is to utilize the intrinsic data correlation to derive self-supervision signals and enhance the data representations via pre-training methods for improving sequential recommendation. Tiansheng Yao et.al \cite{b28} introduced a multi-task self-supervised learning (SSL) framework for large-scale item recommendations by adding regularization to improve generalization. More recently, in \cite{b29}, the authors explored self-supervised learning on user-item graph, so as to improve the accuracy and robustness of GCNs for recommendation.
\section{Proposal model}

\subsection{Architecture}

\begin{figure}[!ht]
\centering \includegraphics[scale=0.32]{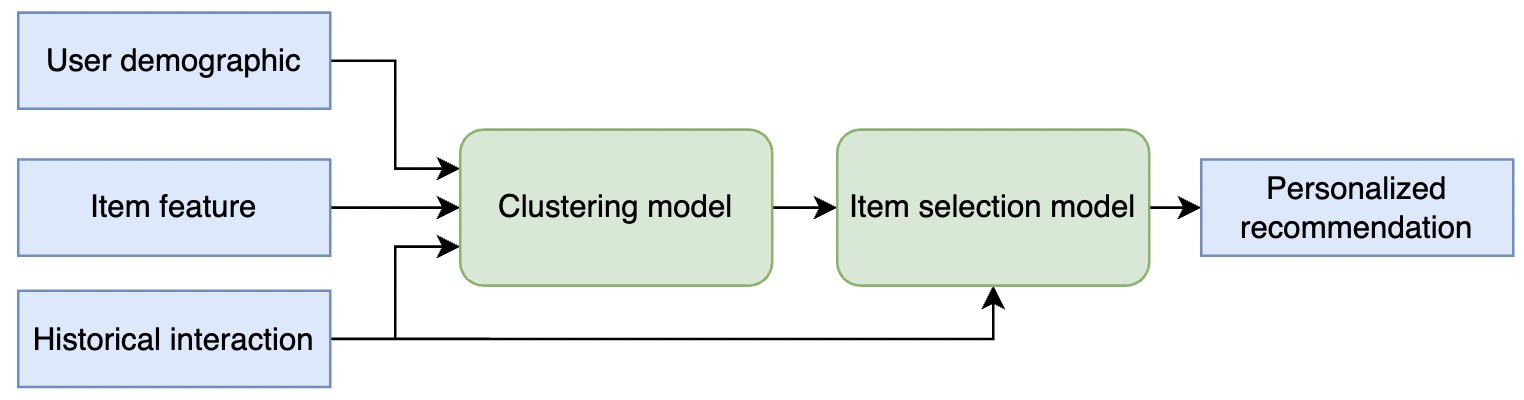}
\caption{Overall architecture}
\label{fig1}
\end{figure}

\par We propose a recommended system architecture consisting of two parts: clustering model and item selection model. The clustering model learns and extracts the characteristics of the user's behavior, and then assigns clusters for them. The clusters identify users with similar preferences. Given the items catalog $ T= \{ t_1, t_2, ..., t_p \}$ with $p$ items. For a sample user $u_i$ with demographic information vector $d_i$, we have a set of interacted items $S_i = \{t_{i1}, t_{i2}, t_{i3},..., t_{iq}; t_{ij} \in T, q \leqslant p \}$. Items in $S_i$ are known the rating value corresponding to the rating vector $r_i = \{r_{i1}, r_{i2}, r_{i3}, ..., r_{iq}\}$. Clustering model $H$ will perform interaction embedding and combine with user demographic $d_i$ to build a vector $z_i$:
$$z_i \leftarrow H (s_i, r_i, d_i)$$
\par $z_i$ is a vector that expresses the user's preference. Using the $z_i$ vector, we identify the top of the most outstanding features of the user behavior and assign clusters by them. With the item selection model, each user cluster that has been identified from the clustering model will be used to select a set of favorite items. For each user, he/she’s clusters will vote to choose the best appropriate items and recommend them for the user.

\subsection{Interaction embedding}
\begin{wrapfigure}{l}[0pt]{0.4\linewidth}
    \includegraphics[scale=0.45]{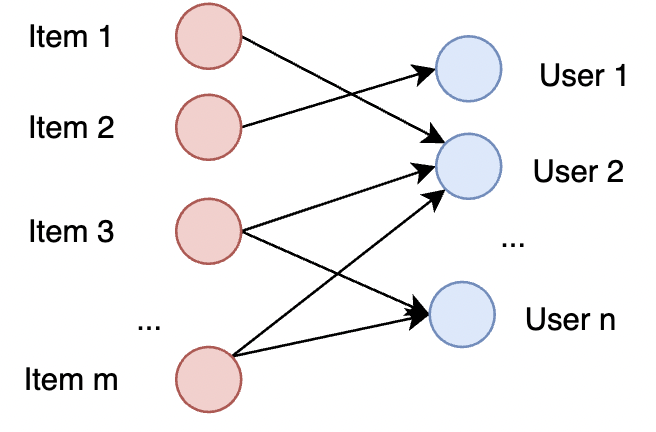}
    \caption{Interactions are considered as a directional graph}
\end{wrapfigure}

\par Similar to \cite{b24}, we consider the interaction data of users and items as a directional graph. In which nodes are users, items and the edges indicate interactive relationships. The magnitude of the edges are rating values. The rating value needs to be normalized in the range of $[-1, 1]$. The positive rating values indicate the user's liking; the negative rating indicates the dislike of the user with the item.
\par Consider the user $u_i$, each interacted item $t_{ij}$ in $S_i$ has two information: feature vector $v_{ij}$ and id of the item $id_{ij}$. The feature vector $v_k$ contains content information of the item such as description, duration, genre, price,... Thus, for the item set $S_i$, we have a matrix of the item features $V_i = \{v_{i1}, v_{i2}, v_{i3}, ..., v_{iq}\}$. The $id_{ij}$ contains hidden information of the item and forms the collaborative filtering properties for the model. We use an embedding layer $EM$ to convert $id_{ij}$ into a representative vector $e_{ij}$: $ e_{ij} \leftarrow EM (id_{ij})$
\par Combine $e_{ij}$ with $v_{ij}$, we obtained a characteristic vector including both hidden and visible information of the item $t_{ij}$: $f_{ij} = concat (v_{ij}, e_{ij})$. Similar to a graph network (GNN), we calculate the $f_{ij}$ for each item in the interactive set $S_i$ and then add them together to obtain the user embedding vector $x_i$. To produce the magnitude of liking/ dislike, we use the rating of each item as an attention module:
$$ x_i  \leftarrow \frac{1}{q}  \sum_{j=1}^{ q } (r_{ij} \times f_{ij})$$


\par We combine $x_i$ with demographic information $d_i$ to obtain the characteristic vector of each user. This vector will go through fully connected layers $F$ to learn and form the preference vector $z_i$ as mentioned in $3.1$: $z_i = F( concat( x_i, d_i) )$
\par
\begin{figure*}[!ht]
  \centering \includegraphics[scale = 0.32]{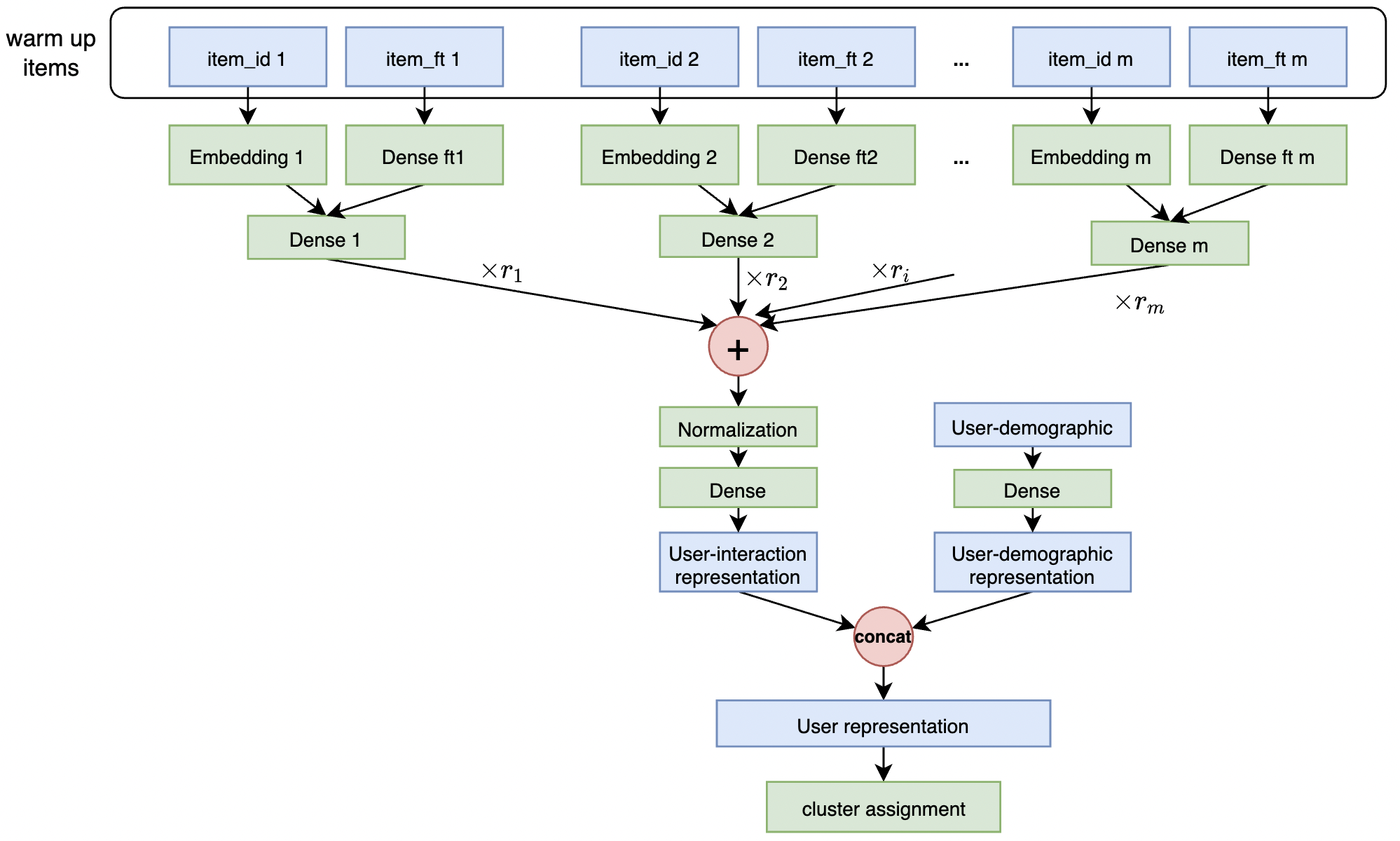}
  \caption{Interaction embedding model.}
  \label{fig4}
\end{figure*}

 The architecture of the preference extracting model shows as Figure \ref{fig4}.

\subsection{Contrastive training}

\par Clustering model $H$ learn a preference vector $z_i \leftarrow H (s_i, r_i, d_i)$ for each user $u_i$. Also, $z_i$ plays the cluster assignment layer. We expect the cluster assignment layer to express the user's preference. Each cluster corresponds to a feature of preference and has a meaningful implication, for example a cluster of action movies, a cluster of long duration movies, a cluster of Jackie Chan’s movies,...or some hidden characteristics that do not exist in feature vector $v_{ij}$ of each item. We expect the following properties of $z_i$:
\begin{itemize}
\item Consistency: Embedding vectors show the preference of a user or very similar users should be close to each other.
\item Distinguishing: Between different users, the embedding vector tends to be different. The more different in user’s preferences, the more distance between embedding vectors.
\end{itemize}

\par We use a triplet contrastive learning architecture \cite{b21} to achieve these goals. For three embedding vector $z_i$, $z^p_i$ which is similar to $z_i$, and $z^n_i$ which is difference with $z_i$, the loss function is define as:
$$L = max( m+\frac{\| z_i -  z^{p}_i \|}{l}-\frac{\| z_i -  z^{n}_i \|}{l}, 0) $$
Where $m$ is the margin and $l$ is the length of the preference vector. 
\par
\begin{figure}[!ht]
\centering \includegraphics[ scale=0.35 ]{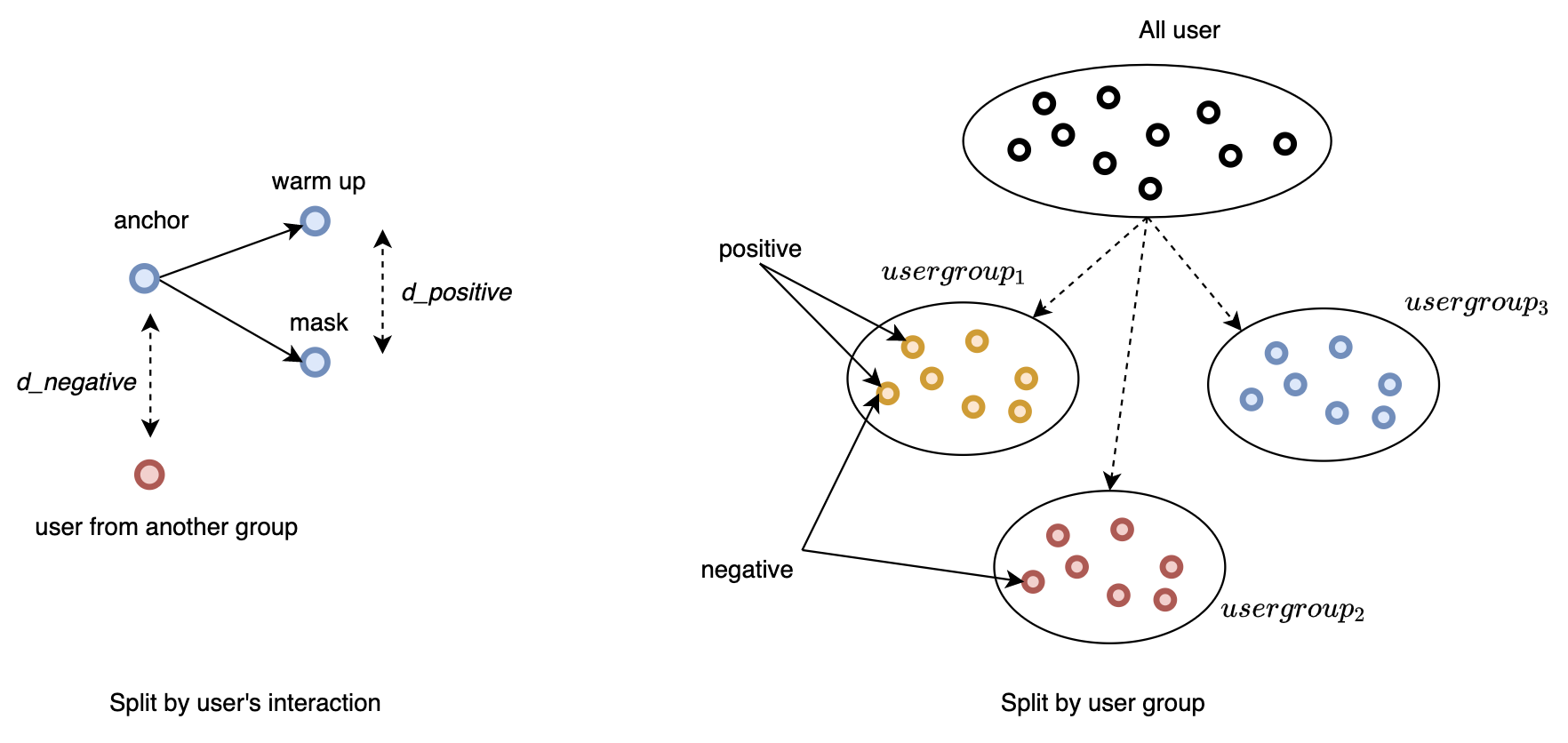}
\caption{Split strategies to get positive and negative representation pairs for triplet contrastive training. This splittings ensure the model leans consistency and distinguishing properties}
\label{fig5}
\end{figure}

 The most important mission now is to choose the right positive and negative pairs. We propose two methods to do that: user’s interaction split and user group split. In the user group split, we categorize users by the movie category they like the most. Then users in the same group are positive with each other and different categories are negative with each other. In the user’s interaction split, the negative pair also are picked users from different groups, but the positive pair are representations from the anchor user and randomly split into two groups: warm up and mask interactions. The reason to choose positive that way is we want the model to generalize the consistency.  The preference vectors extracted from one user should be similar Fig. \ref{fig5}.

\subsection{Item selection pipeline}

\indent\textbf{Soft clustering vs hard clustering}

\par
\begin{figure}[!ht]
\centering \includegraphics[ scale=0.3 ]{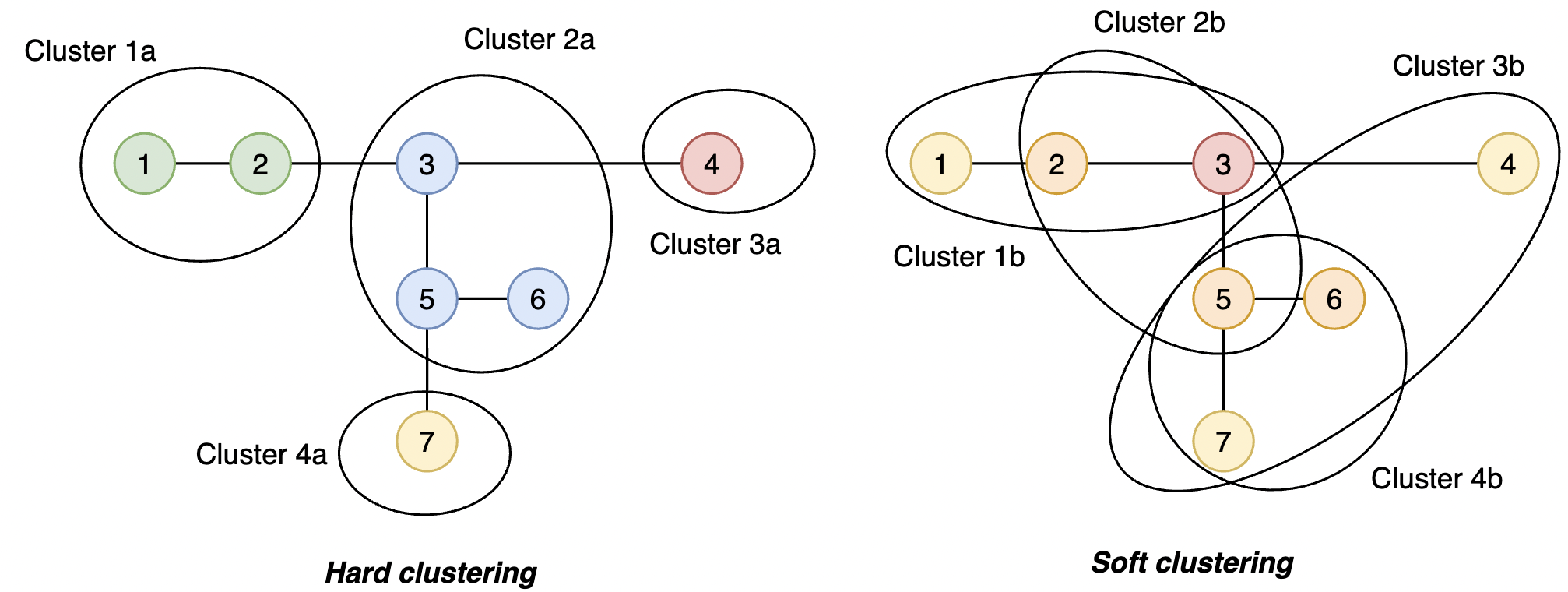}
\caption{Soft clustering compared to hard clustering, for a sparsity dataset, soft clustering lead to more comprehensive view}
\label{fig3}
\end{figure}

 While traditional clustering algorithms try to classify each user (data point) into a cluster such as Kmeans, DBSCAN, ... In a recommender model, this classification have a number of weaknesses:
\begin{itemize}
\item Sparsity: some users are classified alone or are divided into a cluster with very few users. For example, user 7 and user 4 in the Figure \ref{fig3} are classified into clusters 4A and 3A. This makes the number of interactions in the cluster too sparse and does not guarantee the reliability of recommendation.
\item Hard split margin: with points near the classification boundary, although they are very close to each other, they can be classified into 2 completely different clusters. For example, user 5 and user 7, though very close, are divided into 2 cluster clusters 2a, cluster 4a.
\end{itemize}
\par These weaknesses could lead to low efficiency when deploying the recommendation system. We argue that using a soft clustering architecture, in which each data point is allowed to be classified in more than one cluster, would bring better performance in the recommendation field. If a user is divided into a cluster with very few users, then there is still the probability that the user is divided into another cluster with more users. Two users which are close to each other but be splitted into two clusters, can be classified into the same another cluster and thus can still effectively collaborate with each other.

\par To implement soft clustering, we use the user profile vector to be the cluster assignment layer. We use the sigmoid activation function to allow each user to be classified into many different clusters.

\textbf{ Implementation of soft clustering}
\par In order to build a set of recommendations effectively and save computation costs, we will only consider the favorite items. Considering user $u_i$, the preference vector has been calculated as $ z_i = H (S_i) = [rc_{i1}, rc_{i2}, rc_{i3}..rc_{ib}] $, $b$ is the number of clusters. Consider the top $k$ cluster of the user $CL = \argmax(z_i, \text{top\_ k}) = [gr_1, gr_2, gr_3,...gr_k]$ is the top k cluster id best score in $z_i$ corresponding to the confident vector $\zeta \subset z_i$ . Considering the user $u_i$, the favorite set of item is $ S^f_i = [t_{f1}, t_{f2}, ... t_{f \kappa}]$, $S^f_i \subset S_i$, corresponding to rating $r^f_i = [r_{f1}, ... r_{f \kappa}]$ , in which the $t_{fj}$ items are favorited by $u_i$, then $r_{fj}> 0$.

\par We build a shortlist of items by clusters. The shortlist contains the favorite item in accordance with that cluster, the level of confident to determine how the item $g$ is suitable for the cluster $f$ to be determined by:
$$ \Im _{gf} = \frac{\sum_{i=1}^{ N } ( r_{ig} * \zeta_{if})}{\sum_{i=1}^{ N } (\zeta_{if})} $$
\par Where $N$ is total number of user, $r_{ig}$ is rating of user $u_i$ with item $g$, $\zeta_{if}$ is confident score of user $u_i$ to be classified to the cluster $f$. In recommendation phase, for each user, we look up their clusters and let clusters vote for the best suitable items.

 \vspace{0.6cm}

\section{Offline experiment}

\subsection{Experiment setting}

\textbf{ Dataset}

\par We use Movielen 20M (ML20M) \cite{b38},  a relatively large and popular data set in the research field to test our proposed architecture performance. In addition, because ML20M is the dataset without user’s demographic information, we use the Book-crossing dataset (BC) \cite{b39} to examine the effectiveness of embedding demographic information.




\textbf{ Baseline methods}

We use 2 methods that represent two methodologies of implementing the recommendation system:
\begin{itemize}
\item Two-tower DNN (2DNNs)(\cite{b34}, \cite{b35}): Representing the deep learning models that embed information of user and embed information of item, then multiply these two vectors together to create predicted rating. This method requires a large amount of parameters for embedding user\_id and item\_id. Then, the model is trained with the number of observations equal to the observation ratings.
\item Alternating least square (ALS) \cite{b37}: A classic algorithm in the branch of matrix factorization algorithms. This method separates the utility matrix into the hidden matrix of the user and item, then recreates the utility matrix by multiplying these two hidden matrices together.
\end{itemize}

\textbf{ Metrics}
\par As presented at $ 3. $, our architecture will only focus on recommending liked items and ignore unliked items. Therefore, the ranking metrics for such as NDCG, MAP are not suitable to evaluate the recommendation performance. In this study we use precision@50 to evaluate the model performance. The model is considered effective when recommending the products that users like.
$$ precision@k = \frac{ | \Re ^k_i \cap S^f_i | } { | \Re ^k _i \cap S_i | } $$


\subsection{Experiment results}




\par
\begin{figure}[H]
\centering
\begin{subfigure}{.5\textwidth}
  \centering
  \includegraphics[width=1\linewidth, height=6cm]{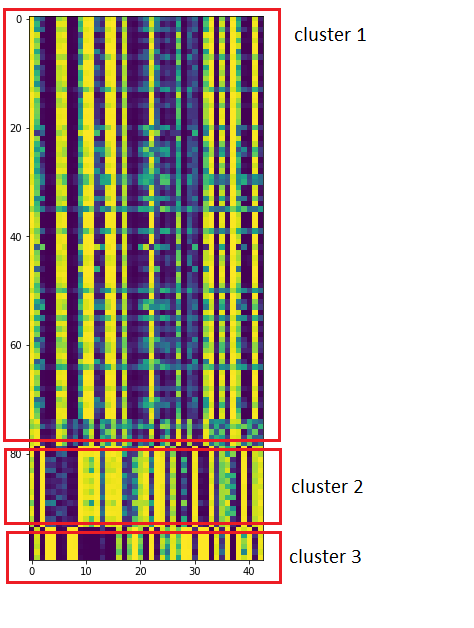}
  \caption{ }
  \label{fig6}
\end{subfigure}%
\begin{subfigure}{.5\textwidth}
  \centering
  \includegraphics[width=1\linewidth, height=6cm]{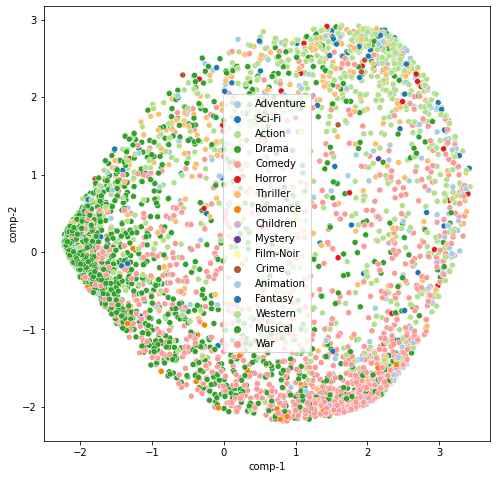}
  \caption{ }
  \label{fig9}
\end{subfigure}
\caption{\small{(A) Spectrogram of user's preference vectors in hard split cluster. Each row is a preference vector of a user. The lighter pixel corresponds with a higher value of the user fitted to the cluster; (B) PCA plots show embedding vectors of user preferences. Labels are the category that users favorite the most. Users with the same first category seem close to each other.}}
\end{figure}

\begin{figure}[H]
\centering
\begin{subfigure}{.5\textwidth}
  \centering
  \includegraphics[width=0.95\linewidth, height=4cm]{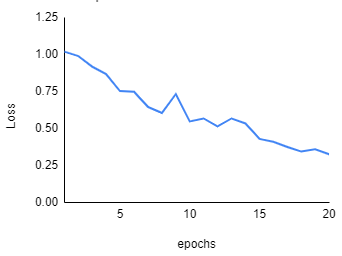}
  \caption{ }
  \label{fig7}
\end{subfigure}%
\begin{subfigure}{.5\textwidth}
  \centering
  \includegraphics[width=0.95\linewidth, height=4cm]{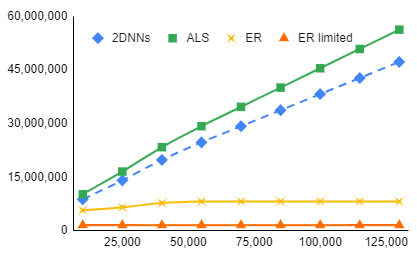}
  \caption{ }
  \label{fig8}
\end{subfigure}
\caption{\small{(A) Loss of user embedding vector model is convergent; (B) Compare models parameters, ER has number of parameter which independent with use’s number and smaller than other models. ER model with limited item catalog has much smaller than full item catalog model}}
\end{figure}

 Training ER model, the loss function tends to decrease after each epoch, proving that the architecture is convergent, the model can learn the characteristics of users based on their interactive history Fig.\ref{fig7} . Figure \ref{fig6} shows the vector preference of some users. After training, the ER model has built a user profile which shows the user's preference and distinguishes the preference of different users. Some users have similar vector profiles and are identified as people who share the same cluster. Figure \ref{fig9} show the embedding vectors by user group. It proves that the contrastive learning has learned the preference of users and can distinguish users with different preferences while pulling users with similar preference to close each other. In addition, the model discovered that users who prefer thriller movies tend to be similar to users who prefer action movies; users that prefer romance movies tend to be close with users that like comedy movies while far from users that like action movies. That discovery makes sense even though no semantic relation between movie categories is provided for the model, then proves the efficiency of the contrastive training.

\begin{table}[!htb]
  \floatsetup{floatrowsep=qquad, captionskip=4pt}
  \begin{floatrow}[2]
    \makegapedcells
    \ttabbox%
    {\begin{tabularx}{0.45\textwidth}{|l| *{2}{>{\centering\arraybackslash}X|}}
      \hline
\textbf{Method}     & \textbf{ML20M} & \textbf{BC} \\ \hline
ALS                  & 6383           & 5189        \\
2DNNs                & 4703           & 4221        \\
ER interaction split & 443            & 382         \\
ER user group split  & 425            & 367         \\ \hline
      \end{tabularx}}
    {\caption[Training time comparition]{Training time comparition}
      \label{val1}}
    \hfill%
    \ttabbox%
    {\begin{tabularx}{0.45\textwidth}{|l| *{2}{>{\centering\arraybackslash}X|}}
     \hline
        \textbf{Method} & {\textbf{ML20M}} & {\textbf{BC}} \\ \hline
        2DNNs                   & 21810                               & 392584                           \\
        ALS                     & 23555                               & 474780                           \\
        ER interaction split             & 516                        & 432                     \\
        ER user group split  & 933            & 839         \\ \hline
      \end{tabularx}}
    {\caption[Inference time comparition]{Inference time comparition}
      \label{val2}}
  \end{floatrow}
\end{table}%

\par The ER model has a significantly lower number of parameters 2DNNs, in addition to 2DNNs and ALS models will have to train with data scores corresponding to the ratings, while the ER training model with the number of data points is corresponding to the user number. In the ML20M data set, the number of users is 145 times less than the number of ratings; in the BC data set, the number of users is 4 times less than the number of user. Combining these two factors helps the ER model training significantly faster 2DNNs. During the training, the ER model runs 12 times faster than 2DNNs with the ML20M data set and is 6 times faster with the BC. While ALS learn linearly correlation of users and items so it has a quite low training time and even quicker than our ER model.

\par In the inference process, while ER only needs to scan and choose on a very limited number of items in the shortlists, the ALS or 2DNNs models must predict for each user pair - the product then sort on the entire catalog item. For each user. This causes ER's inference speed to be 42 times faster than 2DNNs on the ML20M dataset, and on the BC dataset which is a very sparse dataset, ER's Inference speed is even 908 times faster than 2DNNs. These results of ER compared to ALS are even better because ALS gives the computation time longer than 2DNNs.

\begin{table}[htbp]
\caption{Precision@50 comparition between models}
\begin{center}
\setlength{\tabcolsep}{20pt}
\begin{tabular}{lrr}
\hline
\textbf{Method}      & \multicolumn{1}{l}{\textbf{ ML20M}} & \multicolumn{1}{l}{\textbf{ BC  }} \\ \hline
2DNNs                & 0.9721                             & 0.7622                          \\
ALS                  & 0.9582                             & 0.7272                          \\
ER interaction split & 0.9732                             & 0.7736                          \\
ER user group split  & 0.9752                             & 0.7621                          \\ \hline
\end{tabular}
\end{center}
\end{table}


\par Both 2DNNs and ER models are the hybrid of content-based and collaborative filtering while ALS is a pure collaborative filtering model. This gives 2DNNs and ER an advantage compared to ALS at cold start users and sparsity dataset. The test results show that ER for the accuracy of the small-scale recommendation ($k$ = 50 items) is better than the 2DNNs and ALS methods. Especially in the sparse BC dataset ALS gives significantly poor efficiency than ER.

\section{Online Experiments}

\par Since this research is for efficient deployment in industry environment, we conduct an online testing on practical services to evaluate the effectiveness of the model. In the production environment, the data are mainly implicit ratings, observed through the user interactions and sequentially over time. 



\textbf{ Dataset } :
We use TV360 service, with 2 product groups: movies, videos. Movie data has 7000 series of movies, is a relatively dense dataset, while the video product with 200,000 videos is a very sparse data set. With 2 million active users, the scale makes traditional methods extremely expensive. Table .. Compare some statistics of 2 datasets, in which $ sparse ratio = rating\_num / (user\_num \times item\_num)$.

\begin{table}[]
\caption{Statistics of online datasets}
\setlength{\tabcolsep}{10pt}
\begin{tabular}{|l|r|r|}
\hline
\textbf{Dataset info}       & \multicolumn{1}{l|}{\textbf{TV360's films}} & \multicolumn{1}{l|}{\textbf{TV360's videos}} \\ \hline
Active user                 & 1,875,642                                   & 732,514                                      \\ 
Number of item              & 7,251                                       & 185,324                                      \\ 
Number of rating            & 5,347,897                                   & 3,040,837                                    \\ 
Sparse ratio                & 0.0393\%                                     & 0.0022\%                                      \\ 
Liked ratings (implicit)    & 92.01\%                                      & 94.09\%                                       \\ 
Disliked ratings (implicit) & 7.99\%                                       & 5.91\%                                        \\ \hline
\end{tabular}
\end{table}

\textbf{ Metrics} :
We perform A/B testing, by randomly dividing the user volumes into the same 3 sets with the same number of users, homogeneous by stratified sampling. Then we compare the average view number per user (ACPU) and the average watched duration (at second) per user (ADPU).

\textbf{ Results} :
Experimental results show that, for the unbalanced dataset (most interactions are liked) in the industrial environment, the ER model has a significantly better performance than the 2DNNs and ALS models.

\begin{table}[H]
\caption{Results of the online experiments}
\setlength{\tabcolsep}{20pt}
\begin{tabular}{|l|rr|rr|}
\hline
\multirow{2}{*}{\textbf{Methods}} & \multicolumn{2}{l|}{\textbf{TV360's films}}                                                            & \multicolumn{2}{l|}{\textbf{TV360's videos}}                                                           \\ \cline{2-5} 
                                  & \multicolumn{1}{l|}{\textbf{ACPU}} & \multicolumn{1}{l|}{\textbf{ADPU}} & \multicolumn{1}{l|}{\textbf{ACPU}} & \multicolumn{1}{l|}{\textbf{ADPU}} \\ \hline
2DNNs                             & \multicolumn{1}{r|}{0.0152}                 & 13.896                                                   & \multicolumn{1}{r|}{0.0322}                 & 10.526                                                   \\ 
ALS                               & \multicolumn{1}{r|}{0.0121}                 & 12.190                                                   & \multicolumn{1}{r|}{0.0160}                 & 4.277                                                    \\ 
ER interaction split              & \multicolumn{1}{r|}{0.0186}                 & 15.018                                                   & \multicolumn{1}{r|}{0.0421}                 & 12.290                                                   \\
ER user group split               & \multicolumn{1}{r|}{0.0176}                 & 14.272                                                   & \multicolumn{1}{r|}{0.0381}                 & 13.155                                                   \\ \hline
\end{tabular}
\end{table}

\section{Conclusion}
\par Models that their complexity depend on the number of users and items, are extremely expensive in the computation cost, causing great challenges in deploying on the production environment. In this paper, we presented a framework that allows training and inference effectively with low calculation costs and high accuracy. By using interaction embedding, we do not need to use user\_id and allow the model scale to an infinite number of users without affecting the model performance. The calculated data points are each user instead of each rating that help accelerate training significantly, especially for large interactive datasets. Proposals on recommendations according to clustering and shortlist memory gives good effect in recommending and reducing a significant amount of inference costs, especially in sparse dataset. However, clustering recommendations are still an idea that needs more studies to increase accuracy. We expect researchers to continue developing this idea and put it into application in a super -large recommendation system.

\begin{subappendices}
\renewcommand{\thesection}{\Alph{section}}%
\newpage

\section{Online sequencial split}

\par In the online experiment, the ER user interaction split, we divide the data into the corresponding data segments: warm up, mask, target similar to offline experiments, but they are divided sequentially in the sequence of time. Users' interactions show the user's preference, we assume that with movie and video services, this preference is unchanged for less than 6 months. Therefore, the preference vectors of the same user extracted from the warm up, mask set tend to be the same, although different timeframes, the vectors still tend to be similar than preference vectors extracted from different users. Then, we still apply the contrastive learning architecture to train the model. Implicit rating is defined based on the user's viewing time. Because most interactions are reflected when users like the product, to detect the disliked products, we select the products that have been recommended for many times but not be interacted with.

\begin{figure}[!ht]
\centering \includegraphics[ scale=0.6 ]{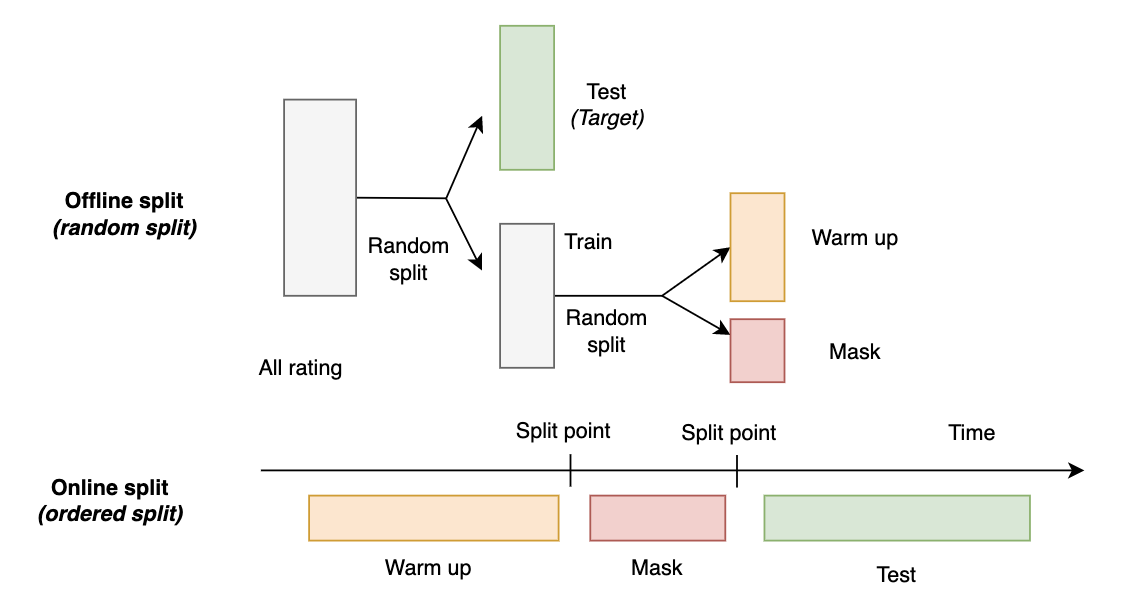}
\caption{Sequencial warm up - mask split versus random split}
\label{fig10}
\end{figure}



\section{Item selection pseudo code}

\par We propose a pseudo code to identify shortlist as  Algorithm\_1. Algorithm\_1 runs independently between users and can be distributed among computers performed in parallel, then summarizing the final $\Im$ according to Map-Reduce methodology.
Shortlist $\Im$ is relatively small because the number of clusters is much smaller than the number of users and items. However, in case of desire to reduce storage capacity, we can sort and retain only the most prominent items for each cluster.
\par Pseudo code to recommend for user $u_i$ after having the shortlist show as Algorithm\_2

    	  \begin{algorithm}[H]
		 \caption{Form shortlist for clusters}
		 \begin{algorithmic}[]
		  \renewcommand{\algorithmicrequire}{\textbf{Input:}}
		 \renewcommand{\algorithmicensure}{\textbf{Output:}}
		 \REQUIRE $H, user\_list, S, r$
		 \ENSURE  shortlist $\Im$
		 \\ \textit{Initialisation} :
		 \\ \STATE $\Im$ = dict\{ cluster\_id: dict\{item : score\}\} ;
		 \\ \textit{// Accumulated rating and counter} :
		\\  \STATE $\Im^{rating}$ = dict\{ cluster\_id: dict\{item:score\}\};  
		 \\ \STATE $\Im^{count}$ = dict\{ cluster\_id: dict\{item: count\}\};

		  \FOR {user\_i in user\_list}
		    \STATE $z_i = H(S_i)$
		    \STATE $S^f_i = S_i $[where rating $>0] = [t_{f1}, t_{f2},... t_{f\kappa}] $
		    \STATE $r^f_i = r_i $[where rating $>0] = [r_{f1}, r_{f2},... r_{f\kappa}] $
		    \STATE $CL = \argmax(z_i, $top\_k$)=[gr_1, gr_2,\dots,gr_k]$
		    \STATE $\zeta = z_i[\text{where index in } CL] = [ rc_{gr1}, rc_{gr2},..., rc_{grk}] $
		  \FOR {cluster in $CL$}
		    \FOR {item   in $S^f_i$}
		      \STATE $\Im ^{rating}[\text{cluster}][\text{item}] += r^f_i[\text{item}]*\zeta[\text{cluster}]$
		      \\
		      \STATE $\Im ^{count} [\text{cluster}][\text{item}] += \zeta[\text{cluster}]$
		    \ENDFOR
		  \ENDFOR
		  \ENDFOR
		  \STATE $\Im $[i][j] = $\Im ^{rating}$ [i][j] / $\Im ^{count}$[i][j]
		 \RETURN $\Im $
		 \end{algorithmic} 
		 \end{algorithm}
		  \begin{algorithm}[H]
		 \caption{Recommend item for user}
		 \begin{algorithmic}[]
		 \renewcommand{\algorithmicrequire}{\textbf{Input:}}
		 \renewcommand{\algorithmicensure}{\textbf{Output:}}
		 \REQUIRE $H, u_i, S_i, \Im$
		 \ENSURE  recommendation list $\Re$
		 \\ \textit{Initialisation} :
		 \\ \STATE $\Re_i$ =  dict\{item : score\} ;
		 \\ \textit{// Accumulated rating and counter} :
		\\  \STATE $\Re^{rating}$ = dict\{item:score\};  
		 \\ \STATE $\Re^{count}$ = dict\{item: count\};
		\\   \STATE $z_i = H(S_i)$
		    \STATE $CL = $argmax($z_i$, top k) $= [gr1, gr2,...grk]$
		    \STATE $\zeta = z_i$[index in $CL$] $= [ rc_{gr1},..., rc_{grk}] $
		  \FOR {cluster in $CL$}
		    \FOR {item in $\Im$[cluster]}
		      \STATE $\Re ^{rating}_i$ [item] +=$\Im$ [cluster][item]*$\zeta[\text{cluster}]$
		    \\ \STATE $\Re ^{count}_i$ [item] += $\zeta[\text{cluster}]$
		    \ENDFOR
		  \ENDFOR

		  \STATE $\Re $[j] = $\Re ^{rating}$ [j] / $\Re ^{count}$[j]
		 \RETURN $\Re $
		 \end{algorithmic} 
		 \end{algorithm}

\end{subappendices}

\end{document}